\documentclass[sigconf]{acmart}

\renewcommand\footnotetextcopyrightpermission[1]{} % removes footnote with conference information in first column
\usepackage{url}
\usepackage{booktabs} % For formal tables
\usepackage{graphicx}

\usepackage{algorithm}
\usepackage[noend]{algpseudocode}
\usepackage{xcolor, soul}

\definecolor{lgray}{rgb}{.90,.90,0.90}
\DeclareRobustCommand{\hlgray}[1]{{\sethlcolor{lgray}\hl{#1}}}

\usepackage{colortbl}

\definecolor{lblue}{rgb}{0.7, 0.7, 1}
\DeclareRobustCommand{\hlblue}[1]{{\sethlcolor{lblue}\hl{#1}}}

\begin{document}

%\setlength{\abovedisplayskip}{0.05pt}
%\setlength{\belowdisplayskip}{0.05pt}
%\setlength{\textfloatsep}{0.15pt}
%\addtolength{\parskip}{-0.15pt}
%\addtolength{\baselineskip}{-0.15pt}
%\setlength{\parskip}{0pt}

\title{TransNets: Learning to Transform for Recommendation}

\author{\textbf{Rose Catherine \hspace{1cm} William Cohen}}
\affiliation{%
  \institution{School of Computer Science \\ Carnegie Mellon University \\
  \texttt{\{rosecatherinek,wcohen\}@cs.cmu.edu}}
}
%\email{rosecatherinek, wcohen@cs.cmu.edu}

\begin{abstract}
Recently, deep learning methods have been shown to improve the performance of recommender systems over traditional methods, especially when review text is available.  For example, a recent model, \textit{DeepCoNN}, uses neural nets to learn one latent representation for the text of all reviews written by a target user, and a second latent representation for the text of all reviews for a target item, and then combines these latent representations to obtain state-of-the-art performance on recommendation tasks.  We show that (unsurprisingly) much of the predictive value of review text comes from reviews of the target user for the target item. We then introduce a way in which this information can be used in recommendation, even when the target user's review for the target item is not available.  Our model, called \textit{TransNets}, extends the \textit{DeepCoNN} model by introducing an additional latent layer representing the target user-target item pair.  We then regularize this layer, at training time, to be similar to another latent representation of the target user's
review of the target item.  We show that TransNets and extensions of it improve substantially over the previous state-of-the-art.
\end{abstract}


\maketitle

\section{Introduction}

Using review text for predicting ratings has been shown to greatly improve the performance of recommender systems \cite{leskovec-hft, topicmf_aaai14, rmr_recsys14}, compared to Collaborative Filtering (CF) techniques that use only  past ratings \cite{netflixkoren2, PMF}. Recent advances in Deep Learning research have made it possible to use Neural Networks in a multitude of domains including recommender systems, with impressive results. Most  neural  recommender models \cite{cdl_kdd15, dcf_cikm15, mvdnn_www15, convmf_recsys16, askgru_recsys16} have focussed on the \textit{content} associated with the user and the item, which are used to construct their latent representations. Content associated with a user could include their demographic information, socioeconomic characteristics, their product preferences and the like. Content linked to an item could include their price, appearance, usability and similar attributes in the case of products, food quality, ambience,  service and wait times in the case of restaurants, or actors, director, genre, and similar metadata in the case of movies.   These representations are then fed into a CF-style architecture or a regression model to make the rating prediction.

Review text, unlike content, is not a property of only the user or only the item; it is a property associated with their joint interaction. In that sense,  it is a \textit{context} \cite{context_recsys08} feature. Only a few neural net models \cite{lmlf_recsys15, cnn_wsdm17, attcnn_mlrec17}  have been proposed to date that use review text for predicting the rating.    Of these, the most  recent model, \textit{Deep Cooperative Neural Networks} (DeepCoNN) \cite{cnn_wsdm17} uses neural nets to learn a latent representation for the user from  the text of all reviews written by her and a second latent representation for the item from the text of all reviews that were written for it, and then combines these two representations in a regression layer to obtain state-of-the-art performance on rating prediction.  However, as we will show, much of the predictive value of review text comes from reviews of the target user for the target item, which can be assumed to be available only at training time, and is not available at test time. In this paper, we  introduce a way in which this information can be used in training the recommender system, such that when the target user's review for the target item is not available at the time of prediction, an approximation for it is generated, which is then used for predicting the rating.  Our model, called \textit{\textbf{Trans}formational Neural \textbf{Net}works} (TransNets), extends the DeepCoNN model by introducing an additional latent layer representing an approximation of the review corresponding to the target user-target item pair.  We then regularize this layer, at training time, to be similar to the latent representation of the actual review written by the target user for the target item.  Our experiments illustrate that TransNets and its  extensions give substantial improvements in  rating prediction. 

The rest of this paper is organized as follows. The proposed model and architecture are discussed in detail in Section \ref{sec_prop}. The experiments and results are discussed in Section \ref{sec_exp}. Section \ref{sec_rel} summarizes  related work, and we conclude in Section \ref{sec_concl}. 

%\vspace*{-4pt}
\section{Proposed Method}
\label{sec_prop}

%\subsection{The basic building block}
\subsection{CNNs to process text}

\begin{figure}
\centering
\includegraphics[bb={150 20 550 350}, clip, width=0.5\textwidth]{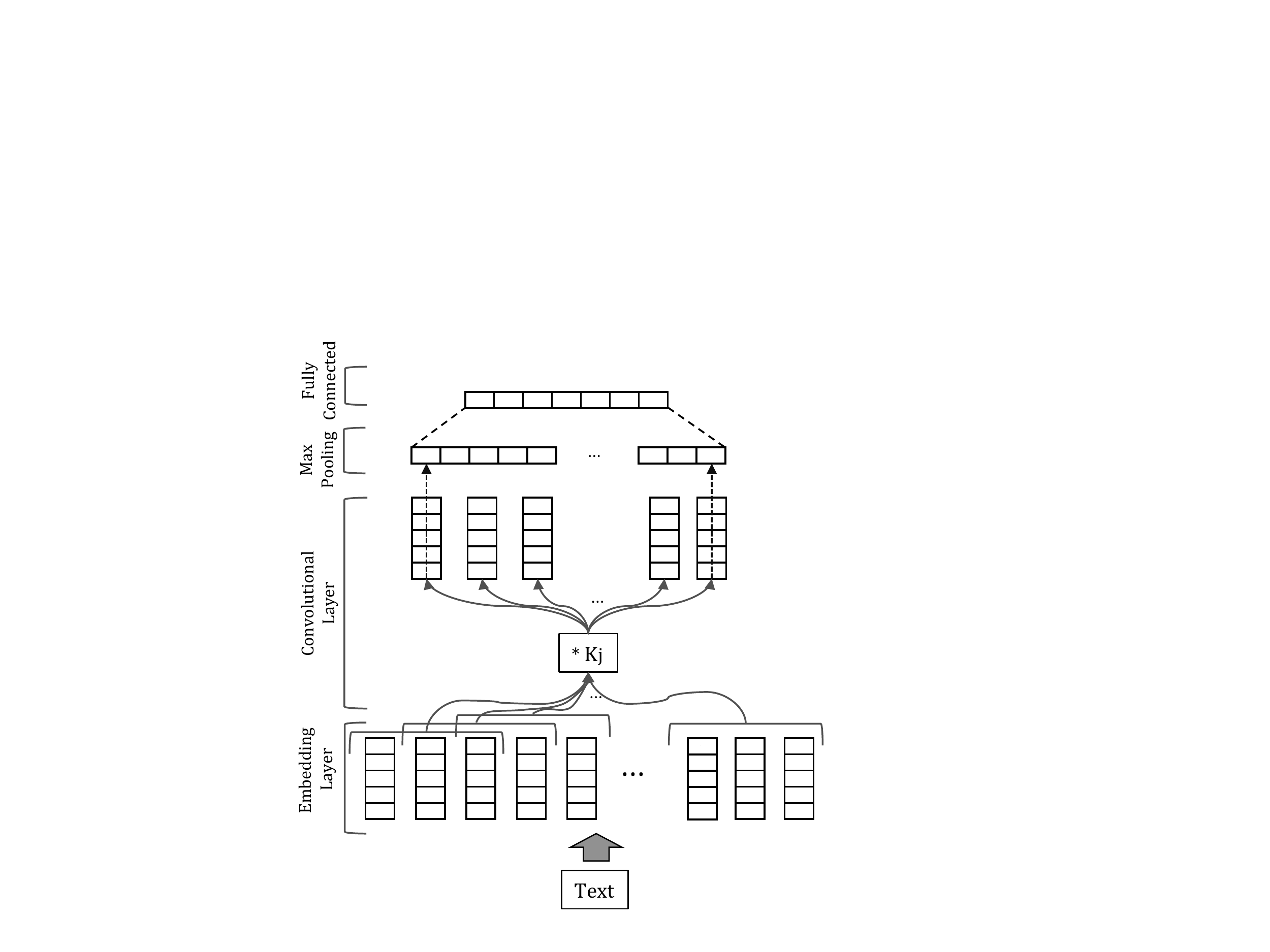}
\caption{The \textit{CNN Text Processor} architecture}
\label{fig_base}
\end{figure}

%As the building block of the approach proposed in this paper, 
We process text using the same approach as  the current state-of-the-art method for rating prediction,  \textit{DeepCoNN}  \cite{cnn_wsdm17}. The basic building block, referred to as a \textit{CNN Text Processor} in the rest of this paper, is a Convolutional Neural Network (CNN) \cite{cnn_ieee98} that inputs a sequence of words and outputs a $n$-dimensional vector representation for the input, i.e., the \textit{CNN Text Processor}  is a function $\Gamma: [w_1, w_2, ..., w_T] \rightarrow \mathbb{R}^n$.  Figure \ref{fig_base} gives the architecture of the \textit{CNN Text Processor}. In the first layer, a word embedding function $f : M \rightarrow \mathbb{R}^d$ maps each word in the  review that are also in its $M$-sized vocabulary into a $d$ dimensional vector. The embedding can be any  pre-trained embedding like those trained on the GoogleNews corpus using \textit{word2vec}\footnote{\url{https://code.google.com/archive/p/word2vec}}\cite{w2v_nips13}, or  on Wikipedia using GloVe\footnote{\url{https://nlp.stanford.edu/projects/glove}} \cite{glove_emnlp14}. These word vectors are held fixed throughout the training process.

Following the embedding layer is the Convolutional Layer, adapted to text processing \cite{nn_nlp_jmlr11}. It consists of $m$ neurons each associated with a filter $K \in \mathbb{R}^{t \times d}$, where $t$ is a window size, typically 2 -- 5. The filter processes $t$-length windows of $d$-dimensional vectors to produce features. Let $V_{1:T}$ be the embedded matrix corresponding to the $T$-length input text. Then, $j^{th}$ neuron produces its features as: 
\begin{eqnarray*}
z_j = \alpha(V_{1:T} * K_j + b_j)
\end{eqnarray*}
where, $b_j$ is its bias, $*$ is the convolution operation and $\alpha$ is a non-linearity like Rectified Linear Unit (ReLU) \cite{relu_icml10} or $tanh$. 

Let $z_j^1, z_j^2, ... z_j^{(T-t+1)}$ be the features produced by the $j^{th}$ neuron on the sliding windows over the embedded text. Then, the final feature corresponding to this neuron is computed using a max-pooling operation, defined as: 
\begin{eqnarray*}
o_j = \max\{z_j^1, z_j^2, ... z_j^{(T-t+1)}\}
\end{eqnarray*}
The max-pooling operation provides location invariance to the neuron, i.e.,  the neuron is able to detect the features  in the text regardless of where it appears. 

The final output of the Convolutional Layer is the concatenation of the output from its $m$ neurons, denoted by:
\begin{eqnarray*}
O = [o_1, o_2, ... o_m]
\end{eqnarray*}

This output is then passed to a fully connected layer consisting of a weight matrix $W \in \mathbb{R}^{m \times n}$ and a bias $g \in \mathbb{R}^n$, which computes the final representation of the input text as: 
\begin{eqnarray*}
x = \alpha(W \times O + g)
\end{eqnarray*}

%\vspace*{-5pt}
\subsection{The DeepCoNN model}

\begin{figure}
\includegraphics[bb={90 80 470 320}, clip, width=0.45\textwidth]{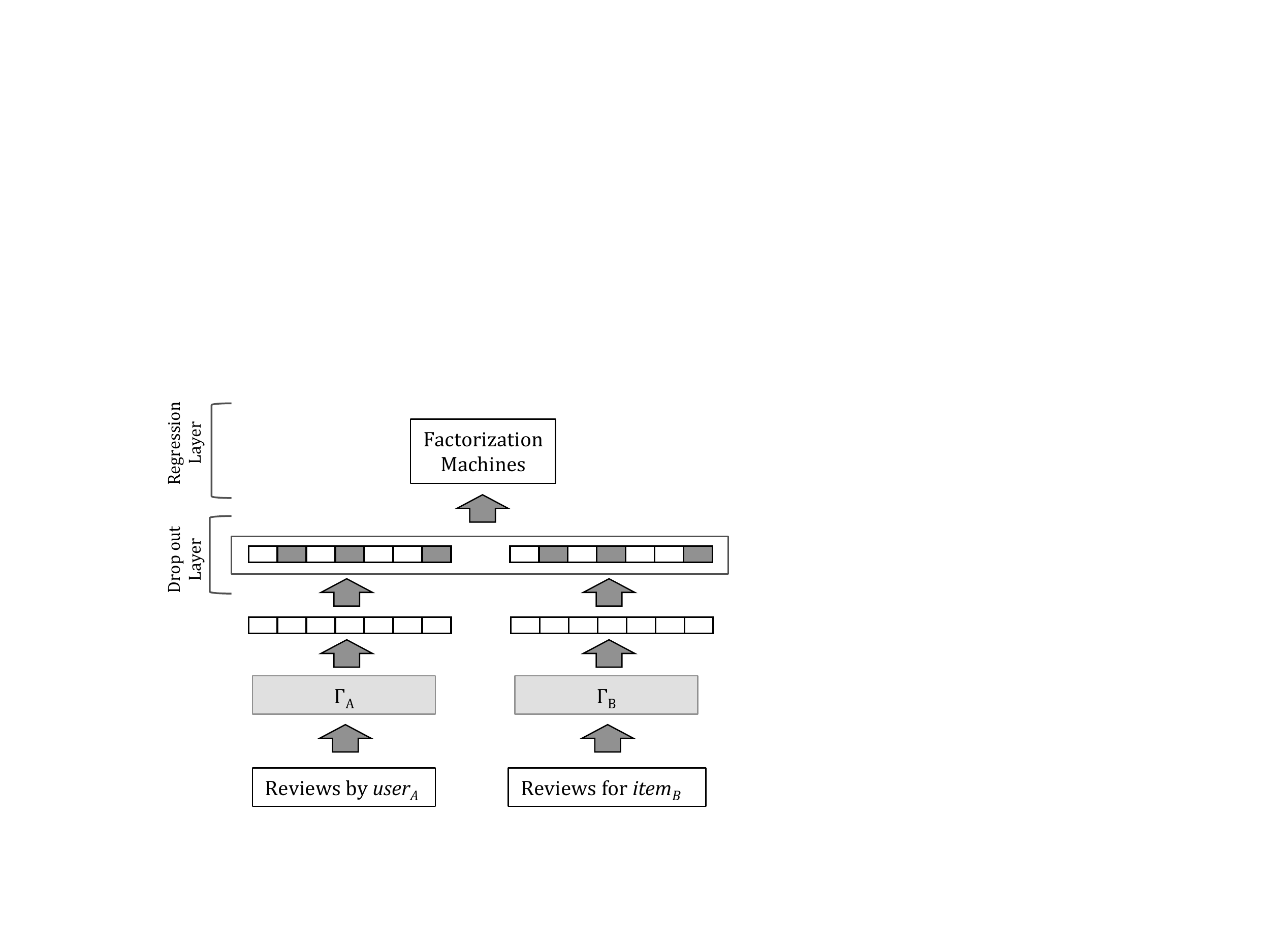}
\caption{DeepCoNN model for predicting rating}
\label{fig_dcnn}
\end{figure}

To compute the rating $r_{AB}$ that  $user_A$ would assign to  $item_B$, the DeepCoNN model of \cite{cnn_wsdm17} uses two \textit{CNN Text Processors}  side by side as  shown in Figure \ref{fig_dcnn}.  The first one processes the text labeled $text_A$, which consists of a concatenation of all the reviews that $user_A$ has written and produces a representation, $x_A$. Similarly, the second  processes the text called  $text_B$, which consists of a concatenation of all the reviews that have been written about  $item_B$ and produces a representation, $y_B$.
Both outputs are passed through a dropout layer \cite{dp_jmlr14}. Dropout is a function $\delta: \mathbb{R}^n \rightarrow \mathbb{R}^n$, that suppresses the output of some of the neurons randomly and is a popular technique  for regularizing a network. Let $\bar{x}_A = \delta(x_A)$ and $\bar{y}_B = \delta(y_B)$, denote the output of the dropout layer applied on $x_A$ and $y_B$. 

The model then concatenates the two representations as $z = [\bar{x}_A, \bar{y}_B]$ and passes it through a regression layer consisting of a \textit{Factorization Machine} (FM) \cite{fm_icdm10}. The FM computes the second order interactions between the elements of the input vector as: 

\begin{eqnarray*}
\hat{r}_{AB} = w_0 + \sum_{i = 1}^{|z|} w_i z_i + \sum_{i = 1}^{|z|} \sum_{j = i+1}^{|z|} \langle\mathbf{v}_i, \mathbf{v}_j \rangle z_i z_j
\end{eqnarray*}
where $w_0 \in \mathbb{R}$ is the global bias, $\mathbf{w}  \in \mathbb{R}^{2n}$ weights each dimension of the input,  and $\mathbf{V} \in \mathbb{R}^{2n \times k}$ assigns a $k$ dimensional vector to each dimension of the input so that the pair-wise interaction between two dimensions $i$ and $j$ can be weighted using the inner product of the corresponding vectors $ \mathbf{v}_i$ and $\mathbf{v}_j$. Note that the FM factorizes the pair-wise interaction, and therefore requires only $O(nk)$ parameters instead of $O(n^2)$ parameters which would have been required otherwise, where $k$ is usually chosen such that $k \ll n$. This has been shown to give better parameter estimates under sparsity \cite{fm_icdm10}. FMs have been used successfully in large scale recommendation services like online news\cite{dwell_recsys14}. 

FMs can be trained using different kinds of loss functions including least squared error ($L_2$), least absolute deviation ($L_1$), hinge loss and logit loss. In our experiments, $L_1$ loss gave a slightly better performance than $L_2$. DeepCoNN \cite{cnn_wsdm17} also uses  $L_1$ loss. Therefore, in  this paper, all FMs are trained using  $L_1$ loss, defined as: 
\begin{eqnarray*}
loss &=& \sum_{(u_A, i_B, r_{AB}) \in D} |r_{AB} - \hat{r}_{AB}|
\end{eqnarray*}

\subsection{Limitations of DeepCoNN}

DeepCoNN model has achieved impressive MSE values surpassing that of the previous state-of-the-art models that use  review texts, like the Hidden Factors as Topics (HFT) model \cite{leskovec-hft},   Collaborative Topic Regression (CTR) \cite{ctr_kdd11} and Collaborative Deep Learning (CDL) \cite{cdl_kdd15}, as well as Collaborative Filtering techniques that use only the rating information like Matrix Factorization (MF) \cite{netflixkoren2} and Probabilistic Matrix Factorization (PMF) \cite{PMF}.

However, it was observed in our experiments that DeepCoNN achieves its best performance only when the text of the review written by the target user for the target item is available at test time. In real world recommendation settings, an item is always recommended to a user \textbf{before} they have experienced it. Therefore, it would be unreasonable to assume that the target review would be available at the time of testing. 

Let  $rev_{AB}$ denote the review written by $user_A$ for an $item_B$. 
At training time,  the text corresponding to $user_A$, denoted as $text_A$, consists of a concatenation of all reviews written by her in the training set. Similarly, the text for $item_B$, denoted by $text_B$, is a concatenation of all reviews written for that item in the training set. Both $text_A$ and $text_B$ includes $rev_{AB}$ for all  $(user_A , item_B)$ pairs in the training set.  At test time, there are two options for constructing the test inputs. For a test pair $(user_{P}, item_{Q})$,  their pairwise review, $rev_{PQ}$ in the test set, could be included in the texts corresponding to the user, $text_P$, and the item, $text_Q$, or could be omitted. In one of our datasets, the MSE obtained by DeepCoNN if $rev_{PQ}$  is included in the test inputs  is only 1.21. However, if $rev_{PQ}$ is omitted, then the performance degrades severely to  1.89. This is  lower than  Matrix Factorization  applied to the same dataset, which has an MSE of  1.86. 
%From such a large disparity in performance, it appears that DeepCoNN may be learning to look for same or similar features appearing in both the representations $x_{A}$ and $y_{B}$ because at training time, $rev_{AB}$ is included in the input, and it will appear in the texts of  both the $user_A$ as well as $item_B$. 
If we train DeepCoNN in the setting that mimics the test setup, by omitting $rev_{AB}$ in the texts of  all $(user_A, item_B)$ pairs in the training set, the performance is  better at 1.70, but still much higher than when  $rev_{AB}$ is available in both training and testing. 

In the setting used in this paper, reviews in the validation and the test set are never accessed at any time, i.e., assumed to be unavailable --- both during training and testing --- simulating a real world situation. 
%However, in certain other tasks like sentiment analysis, $rev_{A'B'}$ can be assumed to be available at test time. 

\subsection{TransNets}

\begin{figure}
\includegraphics[bb={110 75 580 460}, clip, width=0.5\textwidth]{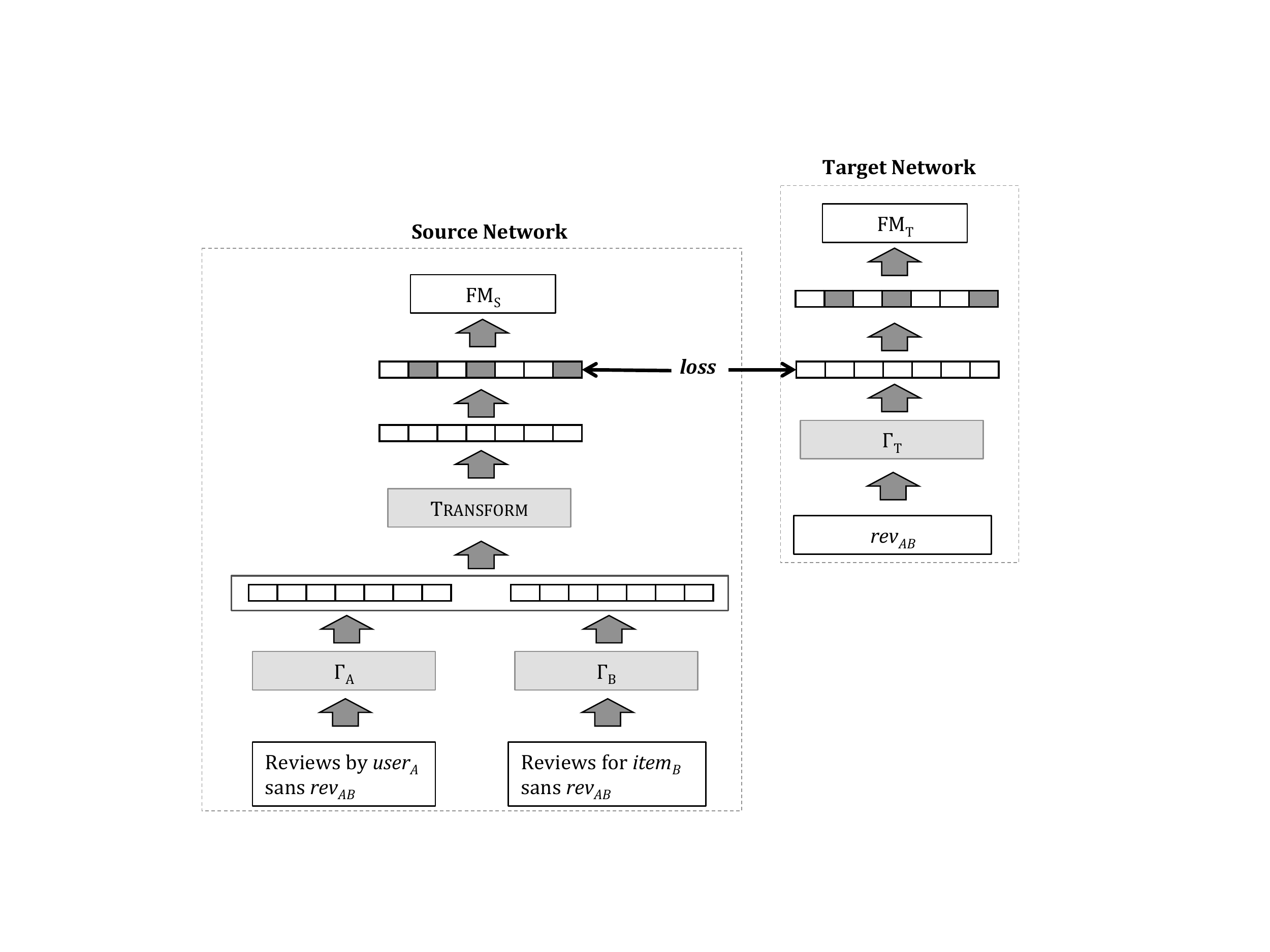}
\caption{The TransNet architecture}
\label{fig_tnn}
\end{figure}

As we saw in the case of DeepCoNN, learning using the target review $rev_{AB}$ at train time inadvertently makes the model dependent on the presence of such reviews at test time, which is unrealistic. 
%of the same review in the texts corresponding to the user and the item. 
However, as shown by the experiment above, $rev_{AB}$  gives an insight into what $user_A$ thought about their experience with $item_B$, and can be an important predictor of the rating $r_{AB}$.  Although unavailable at test time, $rev_{AB}$ is available during training. 

TransNet consists of two networks as shown in the architecture diagram of Figure \ref{fig_tnn}, a \textit{Target Network} that processes the target review $rev_{AB}$ and a \textit{Source Network} that processes the texts of the $(user_A, item_B)$ pair that does not include the joint review, $rev_{AB}$. Given a review text $rev_{AB}$, the Target Network uses a CNN Text Processor, $\Gamma_T$, and  a Factorization Machine, $FM_T$, to predict the rating as:
\begin{eqnarray*}
x_T &=& \Gamma_T(rev_{AB}) \\
\bar{x}_T &=& \delta(x_T) \\
\hat{r}_T &=& FM_T(\bar{x}_T)
\end{eqnarray*}

Since the Target Network uses the actual review, its task is  similar to sentiment analysis \cite{sentiment_emnlp13, sentiment_icml14}. 
%setting is quite like a sentiment analysis framework. 

The Source Network is like the DeepCoNN model with two CNN Text Processors, $\Gamma_A$ for user text, $text_A - rev_{AB}$, and $\Gamma_B$ for item text, $text_B - rev_{AB}$, and a Factorization Machine, $FM_S$, but with an additional \textsc{Transform} layer.  The goal of the \textsc{Transform} layer is to  \textbf{transform} the user and the item texts into an approximation of $rev_{AB}$, denoted by $\hat{rev}_{AB}$, which is then  used for predicting the rating. The Source Network predicts the rating as given below:

First, it converts the input texts into their latent form as: 
\begin{eqnarray*}
x_A &=& \Gamma_A(text_A - rev_{AB}) \\
x_B &=& \Gamma_B(text_B - rev_{AB}) \\
z_0 &=& [x_A x_B]
\end{eqnarray*}
The last step above is a concatenation of the two latent representations. This is then input to the \textsc{Transform} sub-network, which is a $L$-layer deep non-linear transformational network. Each layer $l$ in \textsc{Transform} has a weight matrix $G_l \in \mathbb{R}^{n \times n}$ and bias $g_l \in \mathbb{R}^n$, and transforms its input $z_{l-1}$ as: 
\begin{eqnarray*}
z_{l} &=& \sigma(z_{l-1} G_l + g_l)
\end{eqnarray*}
where $\sigma$ is a non-linear activation function. Since the input to the first layer, $z_0$, is a concatenation of two vectors each of $n$ dimensions, the first layer of \textsc{Transform} uses a weight matrix $G_1 \in \mathbb{R}^{2n \times n}$.

The output of the $L^{th}$ layer of \textsc{Transform}, $z_L$ is the approximation constructed by the TransNet for $rev_{AB}$, denoted by $\hat{rev}_{AB}$. Note that we do not have to generate the surface form of $rev_{AB}$; It suffices to approximate $\Gamma_T(rev_{AB})$, the latent representation of the target review. The Source Network then uses this  representation  to predict the rating as: 
\begin{eqnarray*}
\bar{z}_L &=& \delta(z_L) \\
\hat{r}_S &=& FM_S(\bar{z}_L)
\end{eqnarray*}
During training, we will force the Source Network's representation $z_L$ to be similar to the encoding of $rev_{AB}$ produced by the Target Network, as we discuss below.

%\vspace{-0.85cm}
\subsection{Training TransNets}

% At test time, for a new $user_{P}$ and $item_{Q}$, it can generate  $\hat{rev}_{PQ}$, which can be used as a proxy for $rev_{PQ}$ for predicting the rating. 
%To teach the Source Network how to generate an approximation of the latent representation of the original review $rev_{AB}$ as used by the Target Network, from

TransNet is trained using 3 sub-steps as shown in Algorithm \ref{algo_train}. In the first sub-step, for each training example (or a batch of such examples), the parameters of the Target Network, denoted by $\theta_T$, which includes those of $\Gamma_T$ and $FM_T$, are updated to minimize a $L_1$ loss computed between the actual rating $r_{AB}$ and the rating $\hat{r}_T$ predicted from the actual review text $rev_{AB}$. 

To teach the Source Network how to generate an approximation of the latent representation of the original review $rev_{AB}$ generated by the Target Network, in the second sub-step, its parameters, denoted by $\theta_{trans}$, are updated to minimize a  $L_2$ loss computed between the transformed representation, $\bar{z}_L$, of the texts of the user and the item, and the representation $x_T$ of the actual review. $\theta_{trans}$ includes the parameters of $\Gamma_A$ and $\Gamma_B$, as well as the weights $W_l$ and biases $g_l$ in each of the transformation layers. $\theta_{trans}$ does not include the parameters of $FM_S$.

In the final sub-step, the remaining parameters of the Source Network, $\theta_S$, which consists of the parameters of the $FM_S$ are updated to minimize a $L_1$ loss computed between the actual rating $r_{AB}$ and the rating $\hat{r}_S$ predicted from the transformed representation, $\bar{z}_L$.  
Note that each sub-step is repeated for each training example (or a batch of such examples), and not trained to convergence independently. The training method is detailed in Algorithm \ref{algo_train}.
%Below is the reasoning for some of the design decisions for training TransNets: 
%\begin{itemize}
%\item Training each set of parameters, $\theta_T$, $\theta_{trans}$ and $\theta_S$ to convergence independently vs. 
%\end{itemize}
\begin{algorithm}
\caption{Training TransNet}\label{algo_train}
\begin{algorithmic}[1]
\Procedure{Train}{$D_{train}$}
\While{not converged}
\For{ $(text_A, text_B, rev_{AB}, r_{AB}) \in D_{train}$}
\State \hlgray{\textit{\#Step 1: Train Target Network on the actual review}}
\State $x_T = \Gamma_T(rev_{AB})$
\State $\hat{r}_T = FM_T(\delta(x_T))$
\State $loss_T = |r_{AB} - \hat{r}_T|$
\State update $\theta_T$ to minimize $loss_T$ 
\State \hlgray{\textit{\#Step 2: Learn to Transform}}
\State $x_A = \Gamma_A(text_A)$
\State $x_B = \Gamma_B(text_B)$
\State $z_0 = [x_A x_B]$
\State $z_L = $ \Call{Transform}{$z_0$}
\State $\bar{z}_L = \delta(z_L)$
\State $loss_{trans} = || \bar{z}_L - x_T  ||_2 $
\State update $\theta_{trans}$ to minimize $loss_{trans}$

\State \hlgray{\textit{\#Step 3: Train a predictor on the transformed input}}
\State $\hat{r}_S = FM_S(\bar{z}_L)$
\State $loss_S = |r_{AB} - \hat{r}_S|$
\State update $\theta_S$ to minimize $loss_S$
\EndFor
\EndWhile
\State \Return $\theta_{trans}, \theta_S$
\EndProcedure
\end{algorithmic}
\end{algorithm}

\begin{algorithm}
\caption{Transform the input}\label{algo_trans}
\begin{algorithmic}[1]
\Procedure{Transform}{$z_0$}
\For {layer $l \in L$}
\State $z_l = \sigma(z_{l-1} G_l + g_l)$
\EndFor
\State \Return $z_L$
\EndProcedure
\end{algorithmic}
\end{algorithm}

At test time, TransNet uses only the Source Network to make the prediction  as shown in Algorithm \ref{algo_test}. 
\begin{algorithm}
\caption{Testing  using TransNet}\label{algo_test}
\begin{algorithmic}[1]
\Procedure{Test}{$D_{test}$}
\For{ $(text_P, text_Q) \in D_{test}$}
\State \hlgray{\textit{\#Step 1: Transform the input}}
\State $x_P = \Gamma_A(text_P)$
\State $x_Q = \Gamma_B(text_Q)$
\State $z_0 = [x_P x_Q]$
\State $z_L = $ \Call{Transform}{$z_0$}
\State $\bar{z}_L = \delta(z_L)$
\State \hlgray{\textit{\#Step 2: Predict using the transformed input}}
\State $\hat{r}_{PQ} = FM_{S}(\bar{z}_L)$
\EndFor
\EndProcedure
\end{algorithmic}
\end{algorithm}

\subsection{Design Decisions and Other Architectural Choices}
In this section, we describe some of the choices we have in designing the TransNet architecture and why they did not give good results in our preliminary experiments. 
%\vspace{-4pt}
\subsubsection{\textbf{Training with sub-steps vs. jointly}} \sloppy While training  TransNets using Algorithm \ref{algo_train}, in each iteration (or batch), we could choose to jointly minimize a total loss, $loss_{total} = loss_T \; + \;  loss_{trans} \;  + \;  loss_S$. However, doing so will result in parameter updates to the target network, $\Gamma_T$, resulting from  $loss_{trans}$, in addition to those from $loss_T$, i.e., the Target Network will get penalized for producing a representation that is different from that produced by the Source Network. This results in both networks learning to produce sub-optimal representations and converging to a lower performance in our experiments. Therefore, it is important to separate the Target Network's parameter updates so that it learns to produce the best representation which will enable it to make the most accurate rating predictions from the review text. 
%\vspace{-4pt}
\subsubsection{\textbf{Training Target Network to convergence independently}} We could choose to first train the Target Network to convergence and then train the Source Network to emulate the trained Target Network. However, note that the Target Network's input is the actual review, which is unavailable for testing its performance, i.e., we do not know when the Target Network has converged with good generalization vs. when it is overfitting. The only way to measure the performance at test time is to check the output of the Source Network. Therefore, we let the Source and the Target Networks learn simultaneously and stop when the Source Network's test performance is good. 
%\vspace{-4pt}
\subsubsection{\textbf{Using the same convolutional model to process text in both the Source and Target networks}} We could choose to use the  $\Gamma_T$ that was trained in the Target Network to generate  features from the user and the item text in the Source Network, instead of learning separate $\Gamma_A$ and $\Gamma_B$. After all, we are learning to transform the latter's output into the former.  However, in that case, TransNet would be constrained to generate generic features similar to topics. By providing it with separate feature generators, it can possibly learn to transform the occurrence of different features in the user text and the item text of the Source Network to another feature in the Target Network.  For example, it could learn to transform the occurrence of features corresponding to say, `love indian cuisine' \& `dislike long wait' in the user profile, and `lousy service' \&  `terrible chicken curry' in the item (restaurant) profile, to a feature corresponding to say,  `disappointed' in the target review, and subsequently predict a lower rating. Having separate feature generators in the Source Network gives TransNets more expressive power and gave a better performance compared to an architecture that reuses the Target Network's feature generator. 
\subsubsection{\textbf{Training the \textsc{Transform} without the dropout}} We could choose to match the output $z_L$ of \textsc{Transform} with $x_T$ instead of its dropped out version $\bar{z}_L$ in Step 2 of Algorithm \ref{algo_train}. However, this makes the \textsc{Transform} layer unregularized, leading it to overfit thus giving poor performance.

\subsection{Extended TransNets}

\begin{figure}
\includegraphics[bb={110 95 580 305}, clip, width=0.5\textwidth]{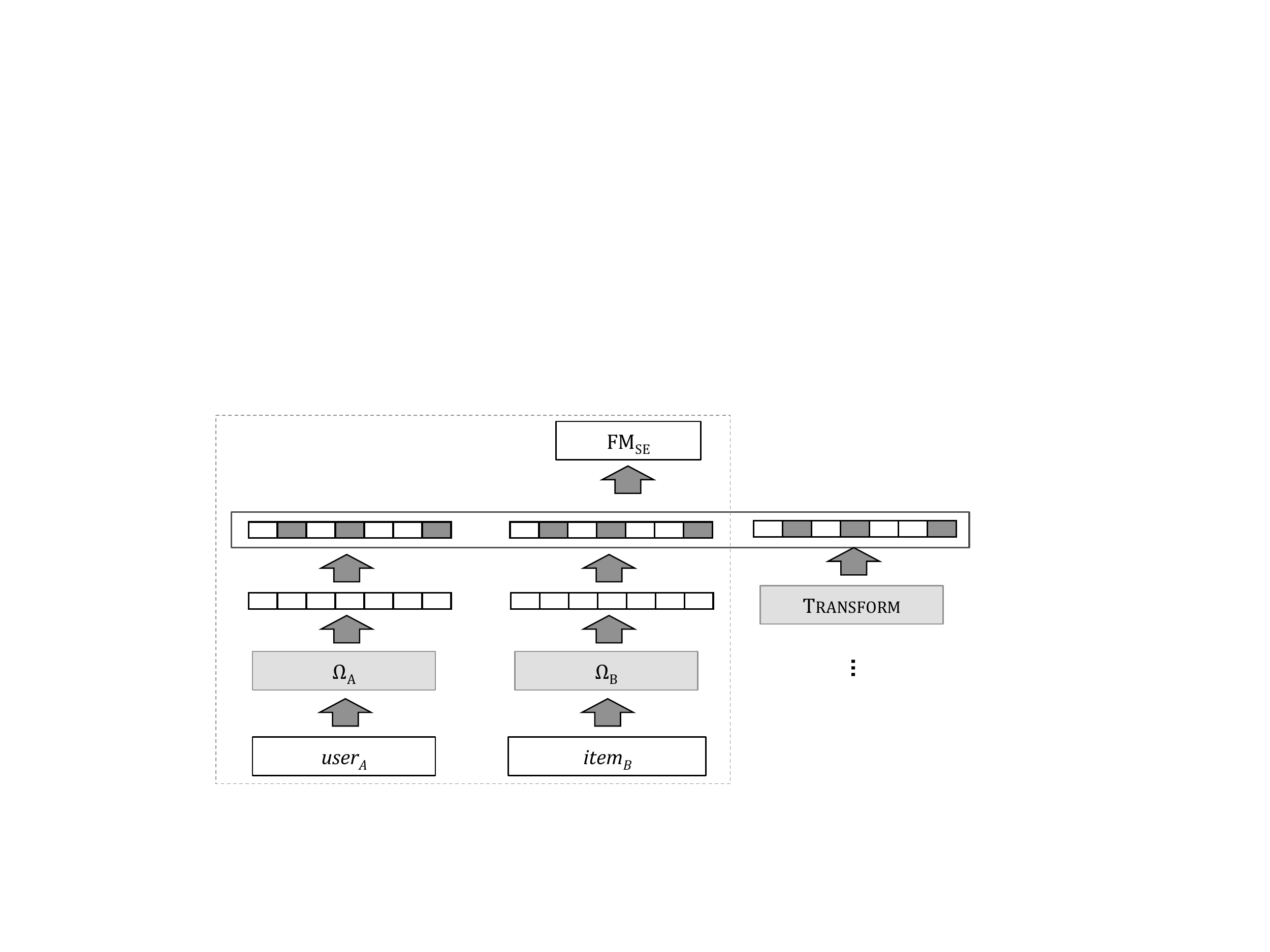}
\caption{The Extended TransNet sub-architecture}
\label{fig_tnne}
\end{figure}

TransNet uses only the text of the reviews and is user/item identity-agnostic, i.e., the user and the item are fully represented using the review texts, and  their identities are not used in the model. However, in most real world settings, the identities of the users and items are known to the recommender system. In such a scenario, it is beneficial to learn a latent representation of  the users and items, similar to Matrix Factorization methods. The \textit{Extended TransNet} (TransNet-Ext) model achieves that by extending the architecture of TransNet as shown in Figure \ref{fig_tnne}. 

%The only change is in $\textit{Step 3}$ of Algorithm \ref{algo_train} where
The Source Network now has two embedding matrices $\Omega_A$ for users and $\Omega_B$ for items, which are functions of the form, $\Omega: id \rightarrow \mathbb{R}^n$. These map the string representing the identity of   $user_A$ and   $item_B$   into a $n$-dimensional representation.  These latent representations are then passed through a dropout layer and concatenated with the output of the \textsc{Transform} layer before being passed to the FM regression layer. Therefore, given $user_A$ and $item_B$, TransNet-Ext computes the rating as: 
\begin{eqnarray*}
\omega_A &=& \Omega(user_A)\\
%\bar{\omega}_A &=& \delta(\Omega(user_A)) \\
\omega_B &=& \Omega(item_B)\\
%\bar{\omega}_B &=& \delta(\Omega(item_B)) \\
\bar{z} &=& [ \delta({\omega}_A) \; \delta({\omega}_B) \;\bar{z}_L] \\
\hat{r}_{SE} &=& FM_{SE}(\bar{z})
\end{eqnarray*}

%The regression layer trained in \textit{Step 3} of Algorithm \ref{algo_train}, $ FM_{SE}$ now uses more parameters since the input is larger. 
Computation of the loss in \textit{Step 3} of Algorithm \ref{algo_train}, $loss_{SE}$ is same as earlier: $loss_{SE} =  |r_{AB} - \hat{r}_{SE}|$. But the parameter $\theta_{S}$ updated at the end 
%of \textit{Step 3} 
now contains the embedding matrices $\Omega_A$ and $\Omega_B$.

\section{Experiments and Results}
\label{sec_exp}

\subsection{Datasets}
We evaluate the performance  of the  approach proposed in this paper on four  large datasets. The first one, \texttt{Yelp17}, is from the latest Yelp  dataset challenge\footnote{\url{https://www.yelp.com/dataset_challenge}},  containing  about 4M reviews and ratings of businesses by about 1M users. The rest are three of the larger datasets in the latest release of Amazon reviews\footnote{\url{http://jmcauley.ucsd.edu/data/amazon}} \cite{az_sigir15, az_kdd15} containing reviews and ratings given by users for products purchased on \texttt{amazon.com}, over the period of May 1996 - July 2014. We use the aggressively de-duplicated version of the dataset and also discard entries where the review text is empty. The statistics of the datasets are given in Table \ref{tab_stats}. The original size of the dataset before discarding empty reviews is given in brackets when applicable.

\begin{table}
  \caption{Dataset Statistics}
  \label{tab_stats}
  \begin{tabular}{l p{1.5cm} ll p{1.5cm}}
    \toprule
    Dataset & Category & \#Users & \#Items & \#Ratings \&  Reviews\\
    \midrule
    \texttt{Yelp17} & & 1,029,432 & 144,072 & 4,153,150 \\
    \midrule
    \texttt{AZ-Elec} & Electronics & 4,200,520  & 475,910  & 7,820,765 (7,824,482) \\
    \texttt{AZ-CSJ} & Clothing, Shoes and Jewelry & 3,116,944  & 1,135,948 & 5,748,260 (5,748,920)\\
    \texttt{AZ-Mov} & Movies and TV & 2,088,428  & 200,915 & 4,606,671 (4,607,047) \\
  \bottomrule
\end{tabular}
\end{table}

\subsection{Evaluation Procedure and Settings}

Each dataset is split randomly into train, validation and test  sets in the ratio $80 : 10 : 10$. After training on every 1000 batches of 500 training examples each, MSE is calculated on the validation and the test datasets. We report the MSE obtained on the test dataset when the MSE on the validation dataset was the lowest, similar to \cite{leskovec-hft}.  
All algorithms, including the competitive baselines, were implemented in Python using TensorFlow\footnote{\url{https://www.tensorflow.org}}, an open source software library for numerical computation, and were trained/tested on NVIDIA GeForce GTX TITAN X GPUs. Training  TransNet on \texttt{Yelp17} takes approximately 40 minutes for 1 epoch ($\sim$6600 batches) on 1 GPU, and gives the best performance in about 2--3 epochs.  

Below are the details of the text processing and the parameter settings used in the experiments:
\vspace{-4pt}
\subsubsection{\textbf{Text Pre-Processing and Embedding}}
All reviews are first passed through a Stanford Core NLP Tokenizer \cite{corenlp} to obtain the tokens, which are then lowercased. Stopwords (\texttt{the}, \texttt{and}, \texttt{is} etc.) as well as punctuations are considered as separate tokens and are retained.  A 64-dimensional  $word2vec$\footnote{\url{https://www.tensorflow.org/tutorials/word2vec\#the_skip-gram_model}} \cite{w2v_nips13}  embedding using the Skip-gram model is pre-trained  on the 50,000 most frequent tokens in each of the training corpora. 
\vspace{-4pt}
\subsubsection{\textbf{CNN Text Processor}}
We reuse most of the hyper-parameter settings reported by the authors of DeepCoNN  \cite{cnn_wsdm17} since varying them did not give any perceivable improvement. In all of the CNN Text Processors $\Gamma_A, \Gamma_B$ and $\Gamma_T$, the number of neurons, $m$, in the convolutional layer  is 100, the window size $t$ is 3, and $n$, the dimension of the output of the CNN Text Processor, is 50. The maximum length of the input text, $T$, is set to 1000. If there are many reviews, they are randomly sorted and concatenated, and the first $T$ tokens of the concatenated version are used.  In our experiments, the word embedding dimension, $d$, is 64, and the vocabulary size, $|M|$ is 50,000. Also, the non-linearity, $\alpha$, is \textit{tanh}. 
\vspace{-4pt}
\subsubsection{\textbf{Dropout Layer and Factorization Machines}} All dropout layers have a keep probability of 0.5.
%
%\subsubsection{\textbf{Factorization Machines}}
In all of the factorization machines, $FM_T, FM_S$ and  $FM_{SE}$, the pair-wise interaction is factorized using a $k = 8$ dimensional matrix, $V$. Since $FM_T$ processes a $n$-dimensional input, its parameters are $\mathbf{w}_T  \in \mathbb{R}^{n}$ and $\mathbf{V}_T \in \mathbb{R}^{n \times k}$. Similarly, since $FM_{SE}$ processes a $3n$-dimensional input, its parameters are $\mathbf{w}_{SE}  \in \mathbb{R}^{3n}$ and $\mathbf{V}_{SE} \in \mathbb{R}^{3n \times k}$. All $\mathbf{w}$'s are initialized to 0.001, and all $\mathbf{V}$'s are initialized from a truncated normal distribution with 0.0 mean and 0.001 standard deviation. All FMs are trained to minimize an $L_1$ loss.
\vspace{-4pt}
\subsubsection{\textbf{\textsc{Transform}}}
The default setting for the number of layers, $L$, is 2. We show the performance for different values of $L$ in Section \ref{sec_L}. All weight matrices $G_l$ are initialized from a truncated normal distribution with 0.0 mean  and 0.1 standard deviation, and all biases $g_l$ are initialized to 0.1. The non-linearity, $\sigma$, is \textit{tanh}.
\vspace{-4pt}
\subsubsection{\textbf{TransNet-Ext}}
The user (item) embedding matrices, $\Omega$, are initialized from a random uniform distribution (-1.0, 1.0), and map users (items) that appear in the training set to a $n = 50$ dimensional space. New users (items) in the validation and test sets are mapped to a random vector. 
\vspace{-4pt}
\subsubsection{\textbf{Training}}
All optimizations are learned using Adam \cite{adam_iclr14}, a stochastic gradient-based optimizer with adaptive estimates,  at a learning rate set to 0.002. All gradients are computed by automatic differentiation in TensorFlow.

\subsection{Competitive Baselines}

We compare our method against the current state-of-the-art, DeepCoNN  \cite{cnn_wsdm17}.  Since DeepCoNN was extensively evaluated against  the previous state-of-the-art models  like Hidden Factors as Topics (HFT) model \cite{leskovec-hft},   Collaborative Topic Regression (CTR) \cite{ctr_kdd11}, Collaborative Deep Learning (CDL) \cite{cdl_kdd15}  and Probabilistic Matrix Factorization (PMF) \cite{PMF}, and shown to surpass their performance by a wide margin, we refrain from repeating those comparisons in this paper. However, we do consider some variations of DeepCoNN. 

Our competitive baselines are:
\begin{enumerate}
\item \textbf{DeepCoNN}: The model proposed in \cite{cnn_wsdm17}. During training, $text_A$ and $text_B$ corresponding to the $user_A$-$item_B$ pair contains their joint review $rev_{AB}$, along with reviews that $user_A$ wrote for other items and what other users wrote for $item_B$ in the training set. During testing, for a $user_P$-$item_Q$ pair,  $text_P$ and $text_Q$ are constructed from only the training set and therefore, does not contain their joint review $rev_{PQ}$.
\item \textbf{DeepCoNN-rev$_{\mathbf{AB}}$}: The same DeepCoNN model (1) above, but trained in a setting that mimics the test setup, i.e., during training, $text_A$ and $text_B$ corresponding to the $user_A$-$item_B$ pair \textbf{does not contain} their joint review $rev_{AB}$, but only the reviews that $user_A$ wrote for other items and what other users wrote for $item_B$ in the training set. Testing procedure is the same as above: for a $user_P$-$item_Q$ pair,  $text_P$ and $text_Q$ are constructed from only the training set and therefore, does not contain their joint review $rev_{PQ}$ which is present in the test set. 
\item \textbf{MF}: A neural net implementation of Matrix Factorization with $n = 50$  latent dimensions. It uses only ratings.
\end{enumerate}

We also provide the performance numbers of DeepCoNN in the setting where the test reviews are available at the time of testing. i.e. the same DeepCoNN model (1) above,  but with the exception that at test time, for a $user_P$-$item_Q$ pair,  $text_P$ and $text_Q$ are constructed from  the training set as well as the test set, and therefore,  contains their joint review $rev_{PQ}$ from the test set.  This is denoted as \textbf{DeepCoNN + Test Reviews}, and its performance is  provided  for the sole purpose of illustrating how much better the algorithm could perform, had it been given access to the test reviews.

\subsection{Evaluation on Rating Prediction}
Like prior work, we use the Mean Square Error (MSE) metric to evaluate the performance of the algorithms. Let $N$ be the total number of datapoints being tested. Then MSE is defined as: 
\begin{eqnarray*}
MSE = \frac{1}{N} \sum_{i = 1}^N (r_i - \hat{r}_i)^2
\end{eqnarray*}
where, $r_i$ is the ground truth rating and $\hat{r}_i$ is the predicted rating for the $i^{th}$ datapoint. Lower MSE indicates better performance.

The MSE values of the various competitive baselines are given in Table \ref{tab_mse}. For each dataset, the best score is highlighted in \colorbox{blue!25}{blue}.  
%For example, in the case of \texttt{Yelp}, the better of DeepCoNN and its variant  DeepCoNN-Test  is   1.7045 obtained by the latter, which is better than  MF (1.8661).

As can be seen from the Table, it is clear that TransNet and its variant TransNet-Ext perform better at rating prediction compared to the competitive baselines on all the datasets (\textit{p-value} $\le 0.05$).  It can also be seen that learning a user and item embedding  using only the ratings in addition to  the  text helps TransNet-Ext  improve the performance over the vanilla TransNet (\textit{p-value} $\le 0.1$), except in the case of one dataset (\texttt{AZ-CSJ}). 

It is also interesting to note that training DeepCoNN mimicking the test setup (DeepCoNN-rev$_{AB}$) 
%vs. DeepCoNN) 
gives a large improvement in the case of \texttt{Yelp}, but does not help in the case of  the \texttt{AZ} datasets.

%\begin{table*}
%  \caption{Performance comparison using MSE metric}
%  \label{tab_mse}
%  \begin{tabular}{l p{1.7cm} c c c c c}
%    \toprule
%    \textbf{Dataset} & \cellcolor{lgray} DeepCoNN + Test Reviews & \textbf{MF} & \textbf{DeepCoNN} & \textbf{DeepCoNN-Test} & \textbf{TransNet} & \textbf{TransNet-Ext} \\
%    \midrule
%   \textbf{ \texttt{Yelp17}} & \cellcolor{lgray} 1.2106 &\cellcolor{red!25} 1.8661 & 1.8984 & 1.7045 & 1.6387 [3.86\%]& \cellcolor{blue!25} 1.5913 [6.64\%] \\
%    \midrule
%    \textbf{\texttt{AZ-Elec}} & \cellcolor{lgray} 0.9791 & 1.8898 & \cellcolor{red!25} 1.9704 & 2.0774 & 1.8380 [2.74\%] & \cellcolor{blue!25}1.7781 [5.91\%] \\
%    \textbf{\texttt{AZ-CSJ}} & \cellcolor{lgray} 0.7747 & 1.5212 & \cellcolor{red!25} 1.5487 & 1.7044 & \cellcolor{blue!25}1.4814 [2.62\%] & 1.5087 [0.82\%] \\
%    \textbf{\texttt{AZ-Mov}} &\cellcolor{lgray}  0.9392 & \cellcolor{red!25} 1.4324 & 1.3611 & 1.5276 & 1.3599 [0.09 \%] & \cellcolor{blue!25} 1.2691 [6.76\%] \\
%  \bottomrule
%\end{tabular}
%\end{table*}

\begin{table*}[h]
  \caption{Performance comparison using MSE metric}
  \label{tab_mse}
  \begin{tabular}{l p{1.7cm} c c c c c}
    \toprule
    \textbf{Dataset} & \cellcolor{lgray} DeepCoNN + Test Reviews & \textbf{MF} & \textbf{DeepCoNN} & \textbf{DeepCoNN-rev$_{\mathbf{AB}}$} & \textbf{TransNet} & \textbf{TransNet-Ext} \\
    \midrule
   \textbf{ \texttt{Yelp17}} & \cellcolor{lgray} 1.2106 &1.8661 & 1.8984 & 1.7045 & 1.6387 & \cellcolor{blue!25} 1.5913  \\
    \midrule
    \textbf{\texttt{AZ-Elec}} & \cellcolor{lgray} 0.9791 & 1.8898 & 1.9704 &   2.0774 & 1.8380  & \cellcolor{blue!25}1.7781  \\
    \textbf{\texttt{AZ-CSJ}} & \cellcolor{lgray} 0.7747 & 1.5212 &  1.5487 & 1.7044 & \cellcolor{blue!25} 1.4487  & 1.4780  \\
    \textbf{\texttt{AZ-Mov}} &\cellcolor{lgray}  0.9392 & 1.4324 & 1.3611 & 1.5276 & 1.3599  & \cellcolor{blue!25} 1.2691 \\
  \bottomrule
\end{tabular}
\end{table*}

\subsection{Picking the number of \textsc{Transform} layers}
\label{sec_L}

\begin{figure}
\includegraphics[width=0.48\textwidth, height=0.24\textwidth]{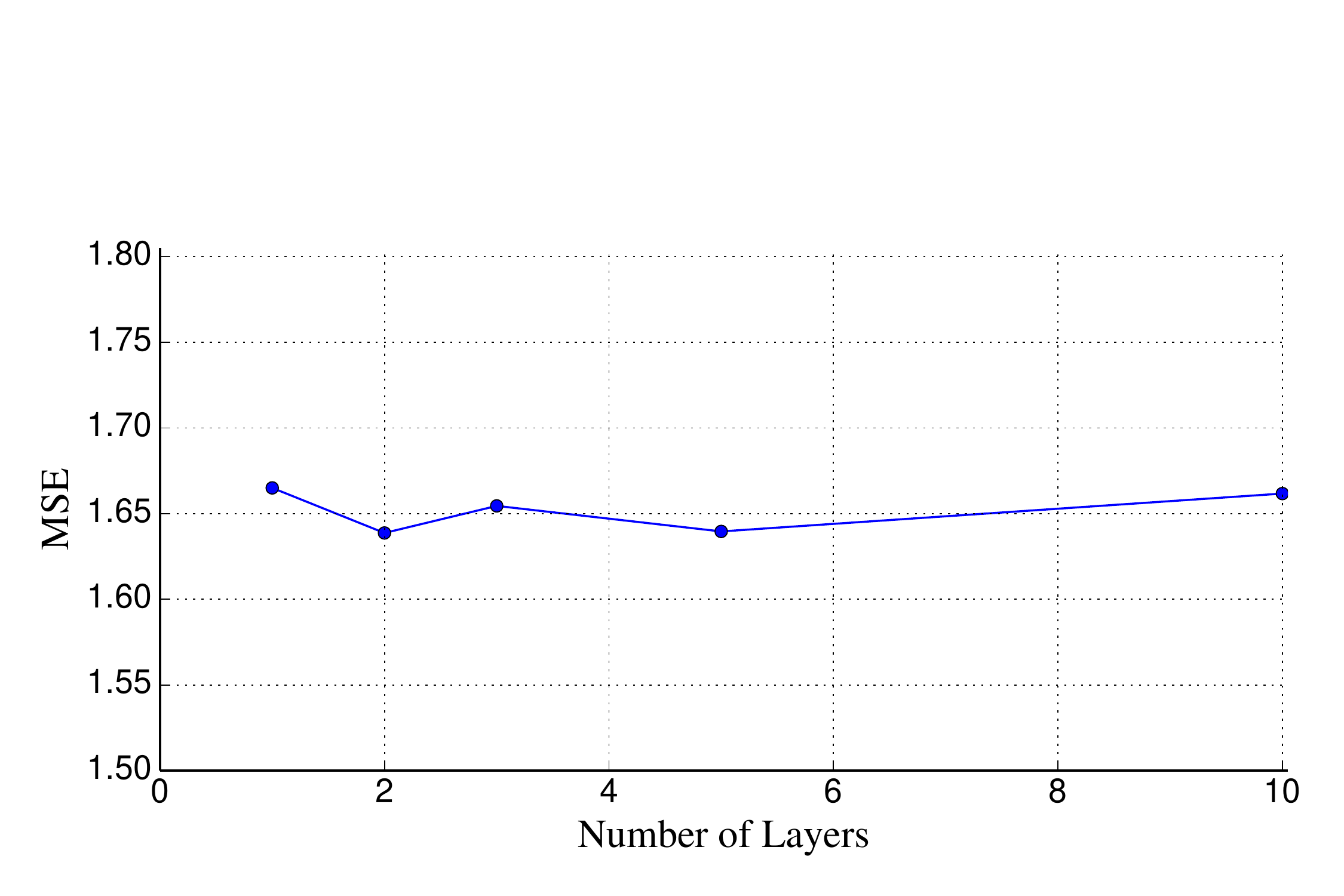}
\caption{Variation in MSE with different Layers: TransNets on \texttt{Yelp} dataset}
\label{fig_layer}
\end{figure}

The \textsc{Transform} network uses $L$  fully connected layers. In Figure \ref{fig_layer}, we plot the  MSE of TransNet on the \texttt{Yelp17} dataset when varying $L$ from 1 to 10. It can be seen from the figure that TransNets are quite robust to the choice of $L$, fluctuating only narrowly in its performance. Using only one layer gives the highest MSE, most probably because it doesn't have enough parameters to learn how to transform the input. Using 10 layers also gives a high MSE, probably because it overfits or because it has too many parameters to learn. From the figure, using 2 or 5 layers gives the best MSE for this particular setting of TransNets. It is known that a 2 layer non-linear neural network is sufficient to represent all the logic gates including the XOR \cite{dlbook}. So, using  2 layers seems like a reasonable choice. 
%The rest of the experiments in this paper use 2 layers in the \textsc{Transform} network. 

%\vspace*{-25pt}
\subsection{Finding the most similar (helpful) reviews}
Our primary evaluation of TransNet is quantitative, using MSE of predicted ratings.  We would also like to investigate whether the learned representation is qualitatively useful---i.e., does it capture interesting high-level properties of the user's review.  One possible use of learning representation would be to give the user information about her predicted reaction to the item that is more detailed than a rating. 
In this section, we show how TransNets could be used to find  reviews that are most similar to what the user would have written, which in turn, could be helpful in making an informed decision. For example, the most helpful review for a user who is more concerned about the quality of service and wait times would be different from the most helpful review for another user who is sensitive to the price. For a test $user_P$-$item_Q$ pair, we run the Source Network with the text of their reviews from the training set to construct $z_L$, which is an approximation for the representation of their actual joint review. Candidate reviews are all the reviews $rev_{CQ}$ in the training set written for $item_Q$ by other users. We pass each of them separately through the Target Network to obtain their latent representation $x_{CQ} = \Gamma_T(rev_{CQ})$. If $rev_{CQ}$ had been most similar to what $user_P$ would write for $item_Q$, then $x_{CQ}$ would be most similar to $z_L$. Therefore, the most similar review is simply the $rev_{CQ}$ whose $x_{CQ}$ is closest to $z_L$ in Euclidean distance.
% has the least squared deviation from $z_L$. 

Some examples of such predicted most similar reviews on the \texttt{Yelp17} dataset are listed in Table \ref{tab_rev_pred}. Here, the column Original Review is the actual review that $user_P$ wrote for $item_Q$, and the column Predicted Review gives the most similar of the candidate reviews predicted by TransNet. The examples show how the predicted reviews talk about particulars that the original reviews also highlight. 

\begin{table*}
\footnotesize
  \caption{Original review vs.  Predicted most helpful review}
  \label{tab_rev_pred}
  \begin{tabular}{p{8.5cm} p{8.5cm} }
  \toprule
    \textbf{Original Review} & \textbf{Predicted Review} \\
    \midrule
my laptop flat lined and i did n't know why , just one day it did n't turn on . i cam here based on the yelp reviews and happy i did . although my laptop could n't be revived due to the fried motherboard , they did give me a full explanation about what they found and my best options . \hlblue{i was grateful they did n't charge me for looking into the problem , other places would have .} i will definitely be coming back if needed . . & my hard drive crashed and i had to buy a new computer . the store where i bought my computer could n't get data off my old hard drive . neither could a tech friend of mine . works could ! \hlblue{they did n't charge for the diagnosis and only charged \$ 100 for the transfer . very happy .} \\
  \midrule
  \hlblue{excellent quality korean restaurant .} it 's a tiny place but never too busy , and \hlblue{quite possibly the best korean dumplings i 've had to date .} & for those who live near by islington station \hlblue{you must visit this new korean restaurant} that just opened up . \hlblue{the food too good to explain . i will just say i havent had a chance to take picture since the food was too grat .} \\
  \midrule
  this place is so cool . the outdoor area is n't as big as the fillmore location , but \hlblue{they make up for it with live music .} i really like the atmosphere and the food is pretty spot on . the sweet potato fry dip is really something special . the vig was highly recommended to me , and i 'm passing that recommendation on to all who read this . & like going on monday 's . happy hour for drinks and apps then at 6pm their burger special . sundays are cool too , when \hlblue{they have live music on their patio .} \\
    \midrule
i have attempted at coming here before but \hlblue{i have never been able to make it in because it 's always so packed with people wanting to eat .} i finally came here at a good time around 6ish ... and not packed but by the time i left , it was packed ! \hlblue{the miso ramen was delicious .} you can choose from add on 's on your soup but they charge you , i dont think they should , they should just treat them as condiments . at other ramen places that i have been too i get the egg , bamboo shoot , fire ball add on 's free . so i am not sure what their deal is . & \hlblue{hands down top three ramen spots on the west coast} , right up there with , and \hlblue{the line can be just as long} . \\
   \midrule
   this place can be a zoo !! however , \hlblue{with the produce they have , at the prices they sell it at , it is worth the hassle .} be prepared to be pushed and shoved . this is much the same as in asia . my wife ( from vietnam ) says that the markets in asia are even more crowded . i agree as i have seen vietnam with my own eyes . & i enjoy going to this market on main street when i am ready to can ... \hlblue{the prices are great} esp for onions . . broccoli and bell peppers ... a few times they have had bananas for \$ 3.00 for a huge box like 30 lbs ... you can freeze them or cover in ... or make banana bread if they begin to go dark ... and ripe . the employees will talk if you say hello first ... \\
   \midrule
   great spot for outdoor seating in the summer since it 's sheltered early from the sun . \hlblue{good service but americanos sometimes are not made right} & this is my `` go to '' starbucks due to the location being close to where i live . i normally go through the drive-thru , which more likely than not , has a long line . . but does n't every starbucks ? i have always received great customer service at this location ! \hlblue{there has been a couple times that my order was n't correct - which is frustrating when you are short on time \& depend on your morning coffee ! but overall you should have a good experience} whether you drive-thru or dine in ! \\
   \midrule
   this is my favorite place to eat in south charlotte . great cajun food . my favorite is the fried oysters with cuke salad and their awesome mac 'n' cheese ( their mac 'n' cheese is not out of a box ) . their sweet tea would make my southern grandma proud . \hlblue{to avoid crowds , for lunch i recommend arriving before 11:30 a.m. or after 1 p.m. and for dinner try to get there before 6 p.m. } is not open on sundays . & always !! \hlblue{too small location so wait line can be long .} been going to for 13 years . \\
   \midrule
   very overpriced food , \hlblue{very substandard food . wait staff is a joke . if you like being ignored in the front room of a one story house and being charged for it , by all means . }otherwise , go to freaking mcdonald 's . & i want this place to be good but it fails me every time . i brought my parents here for dinner and was totally embarrassed with my dining choice . i tried it two more times after that and continue to be disappointed . \hlblue{their menu looks great but what is delivered is a total let down . to top it off , the service is painfully slow .} the only thing this place has going for it is the dog friendly patio and craft beers . \hlblue{i hope someone reads these reviews as the poor service piece continues to be brought up as an issue . } \\
      \midrule
\hlblue{holey moley - these bagels are delicious !} i 'm a bit of a bagel connoisseur . ( note : the bagels at dave 's grocery in ohio city are currently my favs ) . \hlblue{these bagels had me floored .} thankfully , cleveland bagel pops up at festivals and flea markets so there are plenty of opportunities to put them in your mouth ( \hlblue{though rising star coffee is a regular option }) . their are also amazing ! though they are n't the cheapest bagels in the world , you can taste the love that goes into them . they 're perfectly crisp , yet doughy in the middle . the add an added flavor - \hlblue{honestly , it 's a bagel experience .} & i had heard from a colleague at work about cleveland bagel company 's bagels and how they were , `` better than new york city bagels . '' naturally , i laughed at this colleague and thought he was a for even thinking such a thing . so , a few weeks later i happened to be up early on a saturday morning and made the trek up to their storefront -( located across from the harp . ) when i arrived was around 8:15 am ; upon walking in i found most bagel bins to be empty and only a few poppyseed bagels left . i do n't like poppyseed bagels so i asked them what was going on with the rest and when they 'd have more . to my surprise i found out that they only stay open as long as they have bagels to sell . once they sell out , they close up shop and get going for the next day . i ordered a poppyseed bagel even though i do n't like them as i was curious as to what was up with these bagels and \hlblue{can tell you that they are in fact better than new york city bagels . i ca n't even believe i 'm saying that , but it 's true .} you all need to do what you can to get over there to get some of these bagels . \hlblue{they 're unbelievable . i ca n't explain with words exactly why they 're so amazing , but trust me , you will love yourself for eating these bagels . coffee is n't that great , but it does n't matter . get these bagels ?!} \\
\midrule
ok the first time i came here , i was very disappointed in the selection of items , especially after reading previous review . but , then i realized that i went at a bad time , it was the end of the day and they sold out of everything ! \hlblue{i recently went back at the store opening time and a lot happier with the market . they sell freshly made bentos , made in house , and they are perfect for microwaving at home or in the market for a cheap and satisfying meal . the key is to get there early , bc they are limited and run out quick , but they have a good variety of bentos .} one draw back is that it is smaller than expected , so if you come from a place like socal , where japanese markets were like large grocery stores with mini stores and restaurants located inside , you might not be too happy . & \hlblue{the main reason i go here is for the bento boxes} -LRB- see example pic -RRB- . \hlblue{made fresh every day , and when they 're gone , they 're gone . on my way home from work it 's a toss up whether there will be any left when i get there at 5:30 . i would by no means call them spectacular , but they 're good enough that i stop in every weeks} i like to pick up some of the nori maki as well -LRB- see pic -RRB- one thing i wish they had more often is the spam and egg onigiri -LRB- see pic -RRB- . very cool . i 'm told you can order them in advance , so may have to do that \\
  \bottomrule
\end{tabular}
\end{table*}

\section{Related Work}
\label{sec_rel}

%This section presents the relevant prior work, grouped into models for recommendation (Section \ref{sec_rel_model}), and architectures that are comparable to TransNets (Section \ref{sec_rel_arch}). The recommendation models are further grouped into whether they are  Neural Net based or not. 

\subsection{Recommendation Models}
\label{sec_rel_model}
\subsubsection{\textbf{Non-Neural  Models}}
The \textit{Hidden Factors as Topics} (HFT) model  \cite{leskovec-hft} aims to find  topics in the review text that are correlated with the latent parameters of users.  They propose a transformation function which converts user's  latent factors to the topic distribution of the review, and since the former exactly defines the latter, only one of them is learned. 
A  modified version of HFT is the \textit{TopicMF} model \cite{topicmf_aaai14}, where the goal is to match the latent factors learned for the users and items using MF with the topics learned on their joint reviews using a Non-Negative Matrix Factorization, which is then jointly optimized with the rating prediction. In their transformation function, the proportion of a particular topic  in the review is a linear combination of its proportion in  the latent factors of the user and the item, which is then converted into a probability distribution over all topics in that review. Unlike these two models,  TransNet computes each factor in the transformed review from a non-linear combination of any number of factors from the latent representations of either the user or the item or both.
Another extension to HFT is the \textit{Rating Meets Reviews} (RMR) model \cite{rmr_recsys14} where the rating is sampled from a   Gaussian mixture. 
% centered at the latent factor value of the user for that topic. 

The \textit{Collaborative Topic Regression} (CTR) model proposed in \cite{ctr_kdd11} is a content based approach, as opposed to a context / review based approach. It uses LDA \cite{lda_jmlr03} to model the text of documents (scientific  articles), and a combination of MF and content based model for recommendation. 
%For recommending in-matrix documents, they use a combination of matrix factorization and content based approaches. 
%For out-of-matrix recommendations where  no ratings available for a new document, they use the topic proportions of the text of the document to find likely recommendations for the user. 
%
The \textit{Rating-boosted Latent Topics} (RBLT) model of \cite{rblt_ijcai16} uses a simple technique 
%to find recommendable features of products; They repeat
of repeating a review $r$ times in the corpus if it was rated $r$, so that features in higher rated reviews  will dominate the topics. % learned for that item. 
\textit{Explicit Factor Models} (EFM) proposed in \cite{phrase_sigir14} aims to generate explainable recommendations by extracting explicit product features (aspects) and users' sentiments towards these aspects using phrase-level sentiment analysis. 

\subsubsection{\textbf{Neural Net Models}}

The most recent model to successfully employ neural networks at scale for rating prediction  is the \textit{Deep Cooperative Neural Networks} (DeepCoNN) \cite{cnn_wsdm17}, which was discussed in detail in Section \ref{sec_prop}. Prior to that work, \cite{lmlf_recsys15} proposed two models: \textit{Bag-of-Words regularized Latent Factor} model (BoWLF) and \textit{Language Model regularized Latent Factor} model (LMLF), where MF was used to learn the latent factors of users and items, and likelihood of the review text, represented either as a bag-of-words or  an LSTM  embedding \cite{lstm_nc97}, was computed using the  item factors. \cite{attcnn_mlrec17} proposed a CNN based model identical to DeepCoNN, but with  attention mechanism to construct the latent representations, the inner product of which gave the predicted ratings. 

Some of the other past research uses neural networks in a CF setting with content, but not  reviews.  The \textit{Collaborative Deep Learning} (CDL) model \cite{cdl_kdd15} uses a Stacked De-noising Auto Encoder (SDAE) \cite{sdae_jmlr10} to learn robust latent representations of items from their content, which is then fed into a CTR model \cite{ctr_kdd11}  for predicting the ratings. A very similar approach to CDL is the \textit{Deep Collaborative Filtering} (DCF) method \cite{dcf_cikm15} which uses Marginalized De-noising Auto-Encoder (mDA) \cite{mda_icml12} instead.  The \textit{Convolutional Matrix Factorization} (ConvMF) model \cite{convmf_recsys16} uses a CNN  to process the description associated with the item and feed the resulting latent vectors into a PMF  model for  rating prediction.   The \textit{Multi-View Deep Neural Net} (MV-DNN) model \cite{mvdnn_www15} 
%represents users using their search queries and clicked urls, and items (news articles) using their title, named entities in the text and  categories. These features are then processed using a
uses a deep neural net to map  user's  and item's content  into a shared latent space such that their similarity in that space is maximized.  \cite{music_nips13}  proposed to generate the latent factors of items -- music in this case--- from the content, audio signals. The predicted latent factors of the item were then used in a CF style with the  latent factors of the user.   \cite{askgru_recsys16} also proposed a similar technique but adapted to recommending scientific-articles. \cite{youtube_recsys16} used a  deep neural net to learn a latent representation from  video content which is then fed into a deep ranking network.  

Prior research has also used deep neural nets for learning latent factors from ratings  alone, i.e., without using any content or review. \textit{Collaborative De-noising Auto-Encoder} model (CDAE) \cite{cdae_wsdm16}  learns to reconstruct  user's feedback  from a corrupted version of the same. 

\subsection{Comparison to Related Architectures}
\label{sec_rel_arch}
\subsubsection{ Student-Teacher Models}
Student-Teacher models \cite{stud_kdd06, distill_dlw15} also have two networks: a Teacher Network, which is   large and complex, and typically an ensemble of different models, is first trained to make predictions, and a much simpler Student Network, which learns to emulate the output of the Teacher Network, is trained later. 
There are substantial differences between Student-Teacher models and TransNets in how they are structured. Firstly, in Student-Teacher models, the input to both the student and the teacher models are the same. For example, in the case of digit recognition, both networks input the same image of the digit. However, in TransNets, the inputs to the two networks are different. In the Target, there is only one input -- the review  by  $user_A$ for an $item_B$ designated as $rev_{AB}$. But, in the Source, there are two inputs: all the reviews written by $user_A$ sans $rev_{AB}$ and all the reviews written for $item_B$ sans $rev_{AB}$.  Secondly, in Student-Teacher models, the Teacher is considerably complex in terms of width and depth, and the Student is more light-weight, trying to mimic the Teacher. In TransNets, the complexities are reversed. The Target  is lean while the Source  is heavy-weight, often processing large pieces of text using twice the number of parameters as the Target. Thirdly, in Student-Teacher models, the Teacher is pre-trained whereas in TransNets the Target  is trained simultaneously with the Source. A recently proposed Student-Teacher model in \cite{logicnn_2016} does train both the Student and the Teacher simultaneously.  Also, in Student-Teacher models, the emphasis is on learning a simpler and easier model that can achieve similar results as a very complex model. But in TransNets, the objective is to learn how to transform a source representation to a target representation. 

\subsubsection{ Generative Adversarial Networks}
TransNets also bear semblance to GANs \cite{gan_nips14, gan2_icml16} since both are attempting to generate an output which is similar to realistic data. But the models are fundamentally different. Firstly, unlike GAN where  the Generative network generates an output from a random starting point, TransNets have a starting point for each example -- the  reviews written by the user and those written for the item. Secondly, the Adversarial network in GAN tries to classify if the output is real or synthetic. In TransNets, although the objective is to minimize the dissimilarity between the generated representation and that of the real input, there is no  adversarial classifier that attempts to separate each out. Thirdly, in GANs, the adversarial network needs to learn a notion of `real' outputs, which is quite generic. In TransNets, there is always a specific real output to compare to and does not need to learn what a generic real output will look like. 

%\subsubsection{Transfer Learning}
%Transfer Learning is a well known concept in Machine Learning where there are two related tasks -- source and target, and the goal is to improve the performance on the target task by transferring some knowledge from the source task.  A method is a type of Transfer Learning if at least one of the domain or the task is different between the source and the target \cite{tl}. For example, in Transductive Transfer Learning, the source and the target domains are different, but the tasks are the same, while in Inductive Transfer Learning, the source and the target domains are the same, but the tasks are different. In that sense, TransNets is not a type of Transfer Learning because both the domain as well as the task are the same for both the source and the target. 

\section{Conclusions}
\label{sec_concl}

Using reviews for improving recommender systems is an important task and is gaining a lot of attention in the recent years. A recent neural net model, \textit{DeepCoNN}, uses the text of the reviews written by the user and for the item to learn their latent representations, which are then fed into a regression layer for  rating prediction. However, its performance is dependent on having access to the user-item pairwise review, which is unavailable in  real-world settings. 
 
In this paper, we propose a new model called \textit{TransNets} which extends \textit{DeepCoNN} with an additional \textsc{Transform} layer. This additional layer learns to transform the latent representations of user and item into that of their pair-wise review so that at test time, an approximate representation of the target review can be generated and used for making the predictions. We also showed how \textit{TransNets} can be extended to learn user and item representations from ratings only which can be used in addition to the generated review representation. Our experiments showed that  \textit{TransNets} and its extended version can improve the state-of-the-art substantially. 

\begin{acks}
We would like to thank Kathryn Mazaitis for her assistance with the GPU machines. This research was supported in part by Yahoo! through the CMU-Yahoo InMind project.

\end{acks}

\bibliographystyle{ACM-Reference-Format}
\bibliography{mybib} 

\end{document}